\documentclass[11pt]{article}

\usepackage{amsmath,amsthm,amssymb}
\usepackage{fullpage}
\usepackage[parfill]{parskip}
\usepackage[dvipsnames]{xcolor}
\usepackage{libertine}
\usepackage{tikz}
\usetikzlibrary{patterns.meta}
\usepackage{caption}
\usepackage{subcaption}
\usepackage{hyperref}
\usepackage[nameinlink, capitalise]{cleveref}
\usepackage[authoryear]{natbib}
\usepackage{ifthen}

\title{A Geometric Analysis of Gains from Trade}

\author{Jason Hartline\\
    \small Northwestern University\\
    \small Department of Computer Science\\
    \small \href{mailto:hartline@northwestern.edu}{hartline@northwestern.edu} \and
    Kangning Wang\\
    \small Rutgers University\\
    \small Department of Computer Science\\
    \small \href{mailto:kn.w@rutgers.edu}{kn.w@rutgers.edu}}

\date{}

\newtheorem{theorem}{Theorem}
\newcommand{\fb}{\mathtt{FB}}
\renewcommand{\d}{\,\mathrm{d}}
\newcommand{\prob}[2][]{\mathbf{Pr}\ifthenelse{\not\equal{}{#1}}{_{#1}}{}\!\left[{\def\givenn{\middle|}#2}\right]}
\newcommand{\expect}[2][]{\mathbb{E} \ifthenelse{\not\equal{}{#1}}{_{#1}}{}\!\left[{\def\givenn{\middle|}#2}\right]}

\definecolor{linkc}{rgb}{0.1, 0.5, 0.7}
\definecolor{citec}{rgb}{0.4, 0.3, 0.7}
\definecolor{urlc}{rgb}{0.5, 0.1, 0.2}
\hypersetup{
    colorlinks=true,
    linkcolor=linkc,
    citecolor=citec,
    urlcolor=urlc
}

\begin{document}

\maketitle

\begin{abstract}
We provide a geometric proof that the random proposer mechanism is a $4$-approximation to the first-best gains from trade in bilateral exchange. We then refine this geometric analysis to recover the $3.147$-approximation guarantee previously established by \citet{DBLP:conf/wine/Fei22}.
\end{abstract}

\section{Introduction}

\emph{Bilateral trade} is a fundamental economic activity. \citet{myerson1983efficient} considered
the following model: a seller owns a good, and a buyer is interested in purchasing it. The buyer's
value for the good is $v$, and the seller incurs a cost $c$ if the good is sold. The values $v$ and
$c$ are drawn from independent distributions, and both agents have linear utilities.

\citet{myerson1983efficient} studied the social efficiency of mechanisms for bilateral exchanges.
When trade occurs between a buyer with value $v$ and a seller with cost $c$, the resulting social
surplus is $v - c$. The (expected) \emph{gains from trade} of a mechanism refers to the expected
social surplus, where the expectation is taken over the randomness in $v$ and $c$ as well as the
intrinsic randomness of the mechanism. The socially optimal outcome---referred to as the \emph{first
best}---always trades when $v > c$ and never trades when $v < c$. However,
\citet{myerson1983efficient} showed that, for general independent value and cost distributions, no
mechanism has a Bayesian Nash equilibrium (where each agent maximizes expected utility given the
other's strategy) that satisfies individual rationality (so each agent obtains nonnegative utility)
and budget balance (so the mechanism does not require outside subsidy) while achieving the
first-best gains from trade. This result is now known as the \emph{Myerson--Satterthwaite
impossibility theorem}.

This impossibility raises a natural question: is it possible to design an individually rational and
budget-balanced mechanism that, in a Bayesian Nash equilibrium, always achieves a constant-factor
approximation to the first-best gains from trade? Here, for $\alpha \ge 1$, an
$\alpha$-approximation means that the expected gains from trade are at least a $1/\alpha$ fraction
of the first best; smaller ratios are better. \citet*{DBLP:conf/stoc/DengMSW22} answered this
question in the affirmative, showing that the \emph{random proposer} mechanism is an
$8.225$-approximation. For the random proposer mechanism, \citet{DBLP:conf/wine/Fei22} improved the
approximation ratio to $3.147$. More recently, subsequent to our work, the approximation ratio was
improved to $\pi \approx 3.142$ by \citet{jo2026tighter} and $e
\approx 2.718$ by \citet*{bei2026power}, and
\citet*{liu2026secondbest} showed that, for general mechanisms with the aforementioned requirements,
the tight approximation ratio is $2$.

The random proposer mechanism operates as follows: one of the two agents (buyer or seller) is chosen
uniformly at random to act as the proposer. The proposer offers a take-it-or-leave-it price to the
other agent, and the trade occurs if the offer is accepted. In a \emph{perfect Bayesian
equilibrium}, after every offer, the responder accepts when the price gives positive utility and
rejects when it gives negative utility; the proposer chooses an offer maximizing expected utility.

\begin{theorem}[\citealp{DBLP:conf/wine/Fei22}]
For a buyer and a seller with independent values and linear utilities, every perfect Bayesian equilibrium of the random proposer mechanism achieves a $3.147$-approximation to the first-best gains from trade.
\end{theorem}

In this work, we provide a geometric proof of this result. We first establish a simple
$4$-approximation using geometric arguments, and then refine that analysis to recover the
approximation ratio of $3.147$ established by
\citet{DBLP:conf/wine/Fei22}. Our geometric approach is inspired by the simple geometry of auction
equilibria studied by \citet*{DBLP:conf/sigecom/HartlineHT14}, which, as we show, provides new
insights in the context of analyzing bilateral trade. While parts of our analysis follow the
high-level strategy of
\citet*{DBLP:conf/stoc/DengMSW22} and \citet{DBLP:conf/wine/Fei22}, our geometric viewpoint makes
the arguments more intuitive and significantly simpler.

The random proposer mechanism provides a \emph{prior-independent} implementation in perfect Bayesian
equilibrium, i.e., a principal does not need to know the prior distribution to run the mechanism.
While a Bayesian incentive-compatible direct-revelation mechanism achieving the identical bounds
exists via the revelation principle, it would require a \emph{prior-dependent} implementation
\citep[cf.\@][]{DBLP:conf/focs/FengH18}.

For a broader overview of the literature, we refer the reader to the related work section
of \citet*{rubinstein2026multidimensional}.

\section{A Simple \texorpdfstring{$4$}{4}-Approximation}

We first provide a simple proof that the random proposer mechanism is a $4$-approximation to the
first-best gains from trade.

\newcommand{\mathnode}[1]{$#1$}

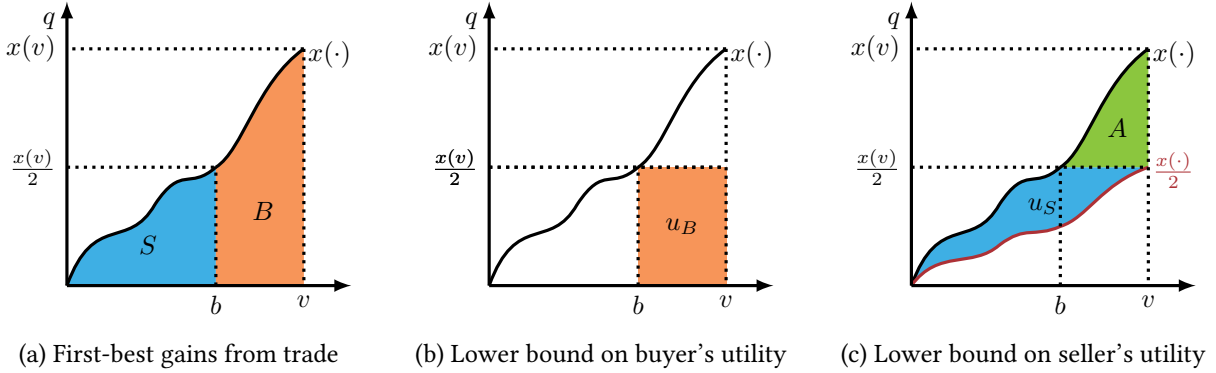
\begin{figure*}[t]
\centering

\tikzset{
    font=\small,
    axis/.style={-latex, very thick},
    maincurve/.style={very thick},
    dashedline/.style={very thick, dotted},
    fillS/.style={fill=CornflowerBlue!90!Black},
    fillB/.style={fill=Peach!75!White}
}

\def\vval{3.5}
\def\xofv{3.5}
\def\bval{2.2}
\def\halfXv{1.75}
\def\figwidth{4.2}
\def\figheight{4.2}

\def\curveSegOne{(0,0) .. controls (0.4, 1.1) and (0.9, 0.5) .. (1.3, 1.2)}
\def\curveSegTwo{(1.3, 1.2) .. controls (1.7, 1.8) and (1.8, 1.4) .. (2.2, 1.75)}
\def\curveSegThree{(2.2, 1.75) .. controls (2.6, 2.0) and (2.8, 3.0) .. (3.5, 3.5)}

\def\curveHSegOne{(0,0/2) .. controls (0.4, 1.1/2) and (0.9, 0.5/2) .. (1.3, 1.2/2)}
\def\curveHSegTwo{(1.3, 1.2/2) .. controls (1.7, 1.8/2) and (1.8, 1.4/2) .. (2.2, 1.75/2)}
\def\curveHSegThree{(2.2, 1.75/2) .. controls (2.6, 2.0/2) and (2.8, 3.0/2) .. (3.5, 3.5/2)}

\def\curveHRSegOne{(1.3, 1.2/2) .. controls (0.9, 0.5/2) and (0.4, 1.1/2) .. (0,0/2)}
\def\curveHRSegTwo{(2.2, 1.75/2) .. controls (1.8, 1.4/2) and (1.7, 1.8/2) .. (1.3, 1.2/2)}
\def\curveHRSegThree{(3.5, 3.5/2) .. controls (2.8, 3.0/2) and (2.6, 2.0/2) .. (2.2, 1.75/2)}

\def\fullCurve{\curveSegOne \curveSegTwo \curveSegThree}

\begin{subfigure}[b]{0.32\textwidth}
    \centering
    \begin{tikzpicture}[scale=0.9]
\def\curvePartS{\curveSegOne \curveSegTwo}
        \def\curvePartB{\curveSegThree}

\fill[fillS] \curvePartS -- (\bval, 0) -- (0,0);
        \fill[fillB] \curvePartB -- (\vval, 0) -- (\bval, 0);
        \node at (1.2, 0.6) {\mathnode{S}};
        \node at (2.9, 1.1) {\mathnode{B}};

\draw[maincurve] \curvePartS \curvePartB node[right, pos=0.95] {\mathnode{x(\cdot)}};

\draw[axis] (0,0) -- (\figwidth,0) node[below left] {\mathnode{}};
        \draw[axis] (0,0) -- (0,\figheight) node[below left] {\mathnode{q}};

\draw[dashedline] (\vval, \xofv) -- (\vval, 0);
        \draw[dashedline] (\vval, \xofv) -- (0, \xofv) node[left] {\mathnode{x(v)}};
        \draw[dashedline] (\bval, 0) -- (\bval, \halfXv);
        \draw[dashedline] (0, \halfXv) node[left] {\mathnode{\frac{x(v)}{2}}} -- (\bval, \halfXv);

\node[below] at (\bval, 0) {\mathnode{b}};
        \node[below] at (\vval, 0) {\mathnode{v}};
    \end{tikzpicture}
    \caption{First-best gains from trade}
    \label{subfig:fb}
\end{subfigure}
\hfill
\begin{subfigure}[b]{0.32\textwidth}
    \centering
    \begin{tikzpicture}[scale=0.9]
\fill[fillB] (\bval, 0) rectangle (\vval, \halfXv);

        \node at ({(\bval+\vval)/2}, {\halfXv/2}) {\mathnode{u_B}};

\draw[maincurve] \fullCurve node[right, pos=0.95] {\mathnode{x(\cdot)}};

\draw[axis] (0,0) -- (\figwidth,0) node[below left] {\mathnode{}};
        \draw[axis] (0,0) -- (0,\figheight) node[below left] {\mathnode{q}};

\draw[dashedline] (\vval, \xofv) -- (\vval, 0);
        \draw[dashedline] (\vval, \xofv) -- (0, \xofv) node[left] {\mathnode{x(v)}};
        \draw[dashedline] (\bval, 0) -- (\bval, \halfXv);
        \draw[dashedline] (0, \halfXv) node[left] {\mathnode{\frac{x(v)}{2}}} -- (\vval, \halfXv);

\node[below] at (\bval, 0) {\mathnode{b}};
        \node[below] at (\vval, 0) {\mathnode{v}};
    \end{tikzpicture}
    \caption{Lower bound on buyer's utility}
    \label{subfig:buyer}
\end{subfigure}
\hfill
\begin{subfigure}[b]{0.32\textwidth}
    \centering
    \begin{tikzpicture}[scale=0.9]
        \def\fullHCurve{\curveHSegOne \curveHSegTwo \curveHSegThree}

\fill[fillS] \curveSegOne -- \curveSegTwo -- (\vval, \halfXv) -- \curveHRSegThree -- \curveHRSegTwo -- \curveHRSegOne -- (0,0);
        \fill[pattern={Lines[angle=35, distance=1.46mm,line width=0.55mm]}, pattern color=LimeGreen!70!Black] \curveSegThree -- (\vval, \xofv) -- (\vval, \halfXv) -- (\bval, \halfXv);

\draw[maincurve] \fullCurve node[right, pos=0.95] {\mathnode{x(\cdot)}};
        \draw[maincurve, color=Maroon, dashed] \fullHCurve node[right, pos=0.95] {\mathnode{\frac{x(\cdot)}{2}}};

        \node at ({\vval/1.8}, {\halfXv/1.5}) {\mathnode{u_S}};
        \node at ({\vval/1.15}, {\xofv/1.5}) {\mathnode{A}};

\draw[axis] (0,0) -- (\figwidth,0) node[below left] {\mathnode{}};
        \draw[axis] (0,0) -- (0,\figheight) node[below left] {\mathnode{q}};

\draw[dashedline] (\vval, \xofv) -- (\vval, 0);
        \draw[dashedline] (\vval, \xofv) -- (0, \xofv) node[left] {\mathnode{x(v)}};
        \draw[dashedline] (\bval, 0) -- (\bval, \halfXv);
        \draw[dashedline] (0, \halfXv) node[left] {\mathnode{\frac{x(v)}{2}}} -- (\vval, \halfXv);

\node[below] at (\bval, 0) {\mathnode{b}};
        \node[below] at (\vval, 0) {\mathnode{v}};
    \end{tikzpicture}
    \caption{Lower bound on seller's utility}
    \label{subfig:seller}
\end{subfigure}

\caption{Geometric analysis of gains from trade}
\label{fig:gains_from_trade}
\end{figure*}

The proof follows from a standard auction-theoretic geometry of the buyer's problem.\footnote{Due to symmetry, focusing on the seller's problem can produce essentially the
same proof.} To set up this geometry, we fix the buyer's value at $v$. (In the end, we will take
the expectation over $v$ to complete the argument.) Let the seller's cost be distributed as follows:
draw a quantile $q \sim U[0,1]$ and define the seller's cost by the non-decreasing function $c(q)$.
(Equivalently, $c(q)$ is the seller's cost at quantile $q$ of the cost distribution.) When the buyer
proposes a price of $b$, let $x(b) = \prob{c(q) \leq b} = c^{-1}(b)$ denote the probability that the
seller accepts.

The first-best gains from trade (conditioned on $v$) is given by the expression
$\expect{\max\bigl(v-c(q),0\bigr)}$, which can be seen geometrically as the colored area in
\Cref{subfig:fb} (by integrating vertically according to the seller's quantile $q$).

Consider the auxiliary price $b = c\bigl(x(v) / 2\bigr)$, i.e., the price at which the seller sells
half as often as when the buyer offers his full value $v$. Geometrically, $b$ divides the first-best
gains from trade into two parts denoted $S$ and $B$ in \Cref{subfig:fb}. This proof will follow the
analysis template from the price of anarchy literature (e.g.,
the work of
\citet*{DBLP:conf/sigecom/HartlineHT14}), where we obtain lower bounds on the agents' utilities
using simple deviation strategies. Specifically, we will obtain a lower bound for the buyer's
utility (when the buyer proposes) that yields at least half of the area $B$, and a lower bound for
the seller's utility (when the seller proposes) that yields at least half of the area $S$.

We can lower bound the buyer's expected utility by the utility he would obtain by offering a price
of $b$. This utility is $u_B = (v - b) \cdot x(b)$, depicted by the orange rectangular area in
\Cref{subfig:buyer}. Since $x(b) = x(v) / 2$, the buyer's utility is at least half of the area $B$
as desired. This inequality also holds in expectation over $v$.

We can lower bound the seller's expected utility by the utility she would obtain by offering a price doubling her quantile, i.e., a price equal to $c\bigl(\min(2q, 1)\bigr)$. As $c(\cdot) = x^{-1}(\cdot)$, the price curve is precisely $x(\cdot) / 2$ when viewed as a function of the horizontal coordinate in \Cref{subfig:seller}. For a fixed quantile $q$, the seller's utility $u_S$ from this offer is
\begin{itemize}
    \item $0$ if the offer is above $v$, and
    \item otherwise, the difference between this offer and the seller's cost, depicted in \Cref{subfig:seller} as the horizontal distance between $x(\cdot)$ and $x(\cdot) / 2$.
\end{itemize}
Taking the expectation over all $q$ is again integrating the vertical axis, which is the solid blue area in \Cref{subfig:seller}. The part of this area that intersects with $S$ is exactly half of $S$, and therefore, the seller's utility using this proposer strategy is at least half of the area $S$ as desired.

For a fixed $v$, the seller's strategy analyzed above does not depend on $v$, and the seller's
utility is at least $S / 2$.\footnote{This step uses the independence of the buyer and seller values. Under
correlation, the definition of $x(\cdot)$ depends on $v$, and then the seller's strategy, which uses
$c(\cdot) = x^{-1}(\cdot)$, also depends on $v$, and hence the argument breaks.} The expected
utility of the seller's optimal strategy, when taking the expectation over $v$, is at least the
seller's expected utility from the analyzed strategy. Thus, taking the expectation over $v$, the
seller's expected utility using her optimal strategy is at least half of the expected area $S$.

To conclude the proof, note that each of the two agents is the proposer half the time; therefore,
the total utility is at least a quarter of the first-best gains from trade.

\section{Recovering Fei's \texorpdfstring{$3.147$}{3.147}-Approximation}

We now refine our geometric proof to recover the approximation ratio of $3.147$ established by
\citet{DBLP:conf/wine/Fei22}. The refinement from $4$ to $3.147$ closely follows the proof
of \citet{DBLP:conf/wine/Fei22}.

As in the geometric proof of the $4$-approximation, we analyze the approximation ratio while fixing
the buyer's value, $v$, and in the end take the expectation over $v$. In \Cref{subfig:seller},
instead of doubling her quantile, the seller now chooses to multiply her quantile by $1 / \lambda$
for a constant parameter $\lambda \in (0, 1)$. The parameter $\lambda$ is used only in the analysis:
each proposer is still chosen with probability $1/2$, independently of $\lambda$. We now use this
constant of $1 / \lambda$ in place of the previous constant of $2$. The dashed red curve in
\Cref{subfig:seller}, in particular, represents the function $\lambda \cdot x(\cdot)$. The figure
shows $\lambda=1/2$; for general $\lambda$, replace $x(\cdot)/2$ and $x(v)/2$ by $\lambda x(\cdot)$
and $\lambda x(v)$, respectively.

Same as before, the first best $\fb$ is the colored area in \Cref{subfig:fb}. The solid blue area
$u_S$ plus the hatched green area $A$ in \Cref{subfig:seller} is exactly $(1 - \lambda) \cdot \fb$.
Here, $u_S$ is a lower bound on the seller's utility when she is the proposer.

Next, we relate the area $A$ with the buyer's utility $u_B$ when he is the proposer. Using the same argument as before, in \Cref{subfig:buyer}, the buyer's utility $u_B$ is lower bounded by the colored rectangular area to the bottom right of any given point on the curve $x(\cdot)$. This means $u_B \geq q \cdot \bigl(v - c(q)\bigr)$ for any quantile $q \in [0, x(v)]$. Using this relation, we can give an upper bound for the area $A$ as follows:
\[
A = \int_{\lambda \cdot x(v)}^{x(v)} \bigl(v - c(q)\bigr) \d q \leq \int_{\lambda \cdot x(v)}^{x(v)} \frac{u_B}{q} \d q = u_B \cdot \ln \frac{1}{\lambda}.
\]

Finally, putting everything together, taking expectations over $v$, and using optimal proposer utilities, we get
\[
(1 - \lambda) \cdot \fb \leq u_S + u_B \cdot \ln \frac{1}{\lambda}.
\]
Here, $u_B$ and $u_S$ now denote expected optimal proposer utilities, and $\fb$ denotes expected first-best gains from trade.
By symmetry, apply the same argument after exchanging the agents and negating their types and the price: buyer value $-c$, seller cost $-v$, and price $-p$. Gains from trade is unchanged, while the two optimal proposer utilities are exchanged, giving
\[
(1 - \lambda) \cdot \fb \leq u_B + u_S \cdot \ln \frac{1}{\lambda}.
\]
In the random proposer mechanism, each agent acts as the proposer with probability $1/2$. Therefore, adding the two inequalities above shows that the mechanism's gains from trade is at least
\[
\frac{u_B+u_S}{2}
\geq\frac{1-\lambda}{1+\ln(1/\lambda)}\fb.
\]

The resulting approximation ratio
\[
\frac{1 + \ln \frac{1}{\lambda}}{1 - \lambda}
\]

is minimized at $\lambda \approx 0.318$, giving a $3.147$-approximation.

\bibliographystyle{plainnat}
\bibliography{refs}

\begin{thebibliography}{9}
\providecommand{\natexlab}[1]{#1}
\providecommand{\url}[1]{\texttt{#1}}
\expandafter\ifx\csname urlstyle\endcsname\relax
  \providecommand{\doi}[1]{doi: #1}\else
  \providecommand{\doi}{doi: \begingroup \urlstyle{rm}\Url}\fi

\bibitem[Bei et~al.(2026)Bei, Wu, and Zhou]{bei2026power}
Xiaohui Bei, Wenhao Wu, and Shengwei Zhou.
\newblock The power of simple mechanisms: A tight {$e$}-approximation for gains
  from trade in matching markets.
\newblock arXiv preprint arXiv:2609.37997, 2026.
\newblock URL \url{https://arxiv.org/abs/2609.37997}.

\bibitem[Deng et~al.(2022)Deng, Mao, Sivan, and Wang]{DBLP:conf/stoc/DengMSW22}
Yuan Deng, Jieming Mao, Balasubramanian Sivan, and Kangning Wang.
\newblock Approximately efficient bilateral trade.
\newblock In \emph{Proceedings of the 54th Annual {ACM} {SIGACT} Symposium on
  Theory of Computing (STOC)}, pages 718--721. {ACM}, 2022.

\bibitem[Fei(2022)]{DBLP:conf/wine/Fei22}
Yumou Fei.
\newblock Improved approximation to first-best gains-from-trade.
\newblock In \emph{Proceedings of the 18th International Conference on Web and
  Internet Economics (WINE)}, pages 204--218. Springer, 2022.

\bibitem[Feng and Hartline(2018)]{DBLP:conf/focs/FengH18}
Yiding Feng and Jason~D. Hartline.
\newblock An end-to-end argument in mechanism design (prior-independent
  auctions for budgeted agents).
\newblock In \emph{Proceedings of the 59th {IEEE} Annual Symposium on
  Foundations of Computer Science (FOCS)}, pages 404--415. {IEEE} Computer
  Society, 2018.

\bibitem[Hartline et~al.(2014)Hartline, Hoy, and
  Taggart]{DBLP:conf/sigecom/HartlineHT14}
Jason~D. Hartline, Darrell Hoy, and Sam Taggart.
\newblock Price of anarchy for auction revenue.
\newblock In \emph{Proceedings of the 15th {ACM} Conference on Economics and
  Computation (EC)}, pages 693--710. {ACM}, 2014.

\bibitem[Jo(2026)]{jo2026tighter}
Sunghyeon Jo.
\newblock Tighter bounds for the random-offerer mechanism in bilateral trade.
\newblock arXiv preprint arXiv:2607.13959, 2026.
\newblock URL \url{https://arxiv.org/abs/2607.13959}.

\bibitem[Liu et~al.(2026)Liu, Qin, Ren, and Wang]{liu2026secondbest}
Zhengyang Liu, Ying Qin, Zeyu Ren, and Zihe Wang.
\newblock Second-best bilateral trade is {$1/2$} efficient.
\newblock arXiv preprint arXiv:2606.03849, 2026.
\newblock URL \url{https://arxiv.org/abs/2606.03849}.

\bibitem[Myerson and Satterthwaite(1983)]{myerson1983efficient}
Roger~B. Myerson and Mark~A. Satterthwaite.
\newblock Efficient mechanisms for bilateral trading.
\newblock \emph{Journal of Economic Theory}, 29\penalty0 (2):\penalty0
  265--281, 1983.

\bibitem[Rubinstein et~al.(2026)Rubinstein, Tan, and
  Zhou]{rubinstein2026multidimensional}
Aviad Rubinstein, Xizhi Tan, and Zixin Zhou.
\newblock Approximately efficient multidimensional bilateral trade.
\newblock In \emph{Proceedings of the 67th {IEEE} Annual Symposium on
  Foundations of Computer Science (FOCS)}. {IEEE} Computer Society, 2026.

\end{thebibliography}

\end{document}